\documentclass[showpacs,twocolumn,prb]{revtex4-1}

\usepackage{amsfonts}
\usepackage{graphicx}
\usepackage{amsmath}
\usepackage{amssymb}

\begin{document}

\title{Optical spectral weight: comparison of weak and strong spin-orbit
coupling }

\author{Zhou Li$^{1,3,4}$}

\email{lizhou@mcmaster.ca}

\author{J. P. Carbotte$^{1,2}$}

\email{carbotte@mcmaster.ca}

\affiliation{$^{1}$ Department of Physics, McMaster University,
Hamilton, Ontario Ontario, Canada,L8S 4M1 \\
 $^{2}$ Canadian Institute for Advanced Research, Toronto, Ontario,
Canada M5G 1Z8 \\
 $^{3}$ Department of Physics and Astronomy, Louisiana State University,
Baton Rouge, LA, 70803 USA \\
 $^{4}$ Center for Computation and Technology, Louisiana State University,
Baton Rouge, LA, 70803 USA} 
\begin{abstract}
The Fermi velocity ($v_{F}$) associated with the spin-orbit coupling
is two orders of magnitude smaller for spintronic semiconductors than
it is for topological insulators. Both families can be treated with
the same Hamiltonian which contains a relativistic (Dirac) linear
in momentum term proportional to $v_{F}$ and a non-relativistic quadratic
contribution with Schr\"{o}dinger mass (m). We find that the AC dynamic
longitudinal and transverse (Hall) magneto-conductivities are strongly
dependent on the size of $v_{F}$. When the Dirac fermi velocity is
small, the absorption background provided by the interband optical
transitions is finite only over a very limited range of photon energies
as compared with topological insulators. Its onset depends on the
value of the chemical potential ($\mu$) and on the magnetic field
(B), as does its upper cut off. Within this limited range its magnitude
is however constant and has the same magnitude of $e^{2}\pi/(8h)$
as is found in topological insulators and also in graphene noting
a difference in degeneracy factor. The total optical spectral weight
under the universal interband background is $e^{2}\pi/(8h)4mv_{F}^{2}$.
In contrast to the known result for graphene no strict conservation
law applies to the spectral weight transfers between inter and intra
band transition brought about by variations in the magnitude of the
chemical potential when a non-relativistic contribution is present
in the Hamiltonian whatever size it may have. 
\end{abstract}

\date{\today }

\maketitle

\section{Introduction}

Since the initial theoretical discussion and experimental discovery
of topological insulators,\cite{Hasan,Qi1,Moore,Hsieh1} their topologically
protected metallic surface states have been extensively studied, both
because of the new physics involved and because of novel functionality
that could find application in a next generation of electronic devices.
Another class of materials which is also of great importance for possible
device applications are those used in studies of spintronics. \cite{Fabian,Zutic,Wolf,DasSarma,Awschalom}Both
classes involve the spin orbit interaction. For topological insulators
this interaction is strong and the linear in momentum relativistic
(Dirac) part of the single particle Hamiltonian proportional to the
Fermi velocity ($v_{F}$) dominates over a smaller non-relativistic
(Schr\"{o}dinger) quadratic in momentum piece characterized by a mass
(m). By contrast, in the materials of interest for spintronics, the
Schr\"{o}dinger contribution dominates over the weak Dirac contribution.
In a first approximation, the same minimal model Hamiltonian can be
used to describe both cases, but the magnitude of the parameters involved
in the Dirac and Schr\"{o}dinger contribution are very different.

In presently studied materials, the Schr\"{o}dinger mass can vary from
the order of the bare electron mass ($m_{e}$) to one tenth its value
and even less. While the Fermi velocity $v_{F}$ is of order $5\times10^{5}$
m/s in topological insulators, in spintronic materials it is less
than a few times $10^{3}$ m/s and often much smaller.\cite{Fabian}
This large difference in the magnitude of the spin orbit coupling
between these two classes of materials can lead to profound differences
in their physical properties. For example in topological insulators
the DC quantum Hall effect shows a quantization $n=1/2,3/2,5/2...$
while in semiconductors it is $n=0,1,2,3...$\cite{Haldane,Gusynin1,Gusynin2,Novo,Zhang,Liu,Li1}

The AC magneto response of materials has been widely studied and provides
valuable information on electron dynamics. Recent related studies
include graphene,\cite{Carbotte1,Carbotte2,Pound,Sadowski,Jiang,Deacon,Onlita}
silicene,\cite{Tabert1,Tabert2} topological insulators,\cite{Li2,Tse,Schafgans}
$MoS_{2}$ \cite{Rose}and Weyl semimetals.\cite{Ashby} In topological
insulators one is dealing with real electron spin and spin momentum
locking has been observed.\cite{Chen1,Xu,Chen2,Hsieh2} For graphene
and other two dimensional membranes such as silicene which displays
buckling and planar $MoS_{2}$, it is the pseudo spin associated with
the two sublattices of the honeycomb lattice which is involved. Weyl
or Dirac semimetals are a three dimensional version of graphene. There 
have also been studies of Kerr and Faraday effects in thin films with 
the breaking of time-reversal symmetry \cite{Tse1,Tse2}. Magneto transport studies \cite{Tkachov1,Tkachov2} 
with emphasis on the transition between ordinary and topological insulator states 
described by a massive Dirac fermion model with change in sign of the mass term. 
Studies are also availale on magneto-optics of bilayer \cite{Abergel}, multilayer \cite{Koshino} graphene 
and graphene on polar substrates \cite{Scharf}. Finally we mention magneto-phonon resonances 
in graphene studied by Raman spectroscopy \cite{Kim}.

In this paper we study the magneto-optical response of a two dimensional
electron gas with Hamiltonian ($H$) consisting of a combination of
relativistic and non-relativistic piece. We feature prominently the
effect of a small spin orbit coupling when the non-relativistic part
of $H$ is dominant. We also compare with results, some known but
many new, that apply in the opposite limit of a topological insulator
for which the spin orbit coupling dominates. Both dynamic AC transverse
($\sigma_{xy}(\omega)$) and longitudinal ($\sigma_{xx}(\omega)$)
conductivity are considered. In section II we provide the formal expressions
for the magneto-optical conductivity based on our chosen Hamiltonian.
We also give simplified but approximate formulas for the optical spectral
weight of the various absorption lines which correspond to the underlying
Landau level (LL) structure created by the magnetic field in the limit
where the Schr\"{o}dinger part is dominant. We work to leading order of
Dirac ($E_{1}$) to Schr\"{o}dinger ($E_{0}$) magnetic energy scale assuming
their ratio to be small. We explicitly consider the case when the
chemical potential $\mu$ is much larger than $E_{0}$. In section
III we provide numerical results for both $Re\sigma_{xx}(\omega)$
and $Im\sigma_{xy}(\omega)$ when $E_{1}$ is increased and is no
longer much smaller than $E_{0}$. Particular attention is paid to
the emergence of the interband background associated with optical
transition between the split helical bands caused by the spin-orbit
coupling. This background is found to be of constant magnitude independent
of the size of $v_{F}$ but is non-zero only in a very limited photon
window. We also present results when the magnetic field is zero. These
greatly help in the physical understanding of the finite $B$ case.
Readjustments of optical spectral weight due to variations in chemical
potential are described. In section IV we turn to the limit when it
is the Dirac energy which dominates. We provide simplified but analytic
expressions which show the first non-zero corrections to the pure
relativistic case for the optical spectral weight when a small subdominant
Schr\"{o}dinger contribution is also included i.e. $E_{0}/E_{1}\ll1$.
Numerical results are provided when our simplified expression is no
longer valid and comparison with the limit $E_{1}/E_{0}\ll1$ is made.
In section V we give more details on spectral weight redistribution
with variations in chemical potential $\mu$. A summary and conclusions
are found in section VI.

\section{Formalism}

\begin{figure}[tp]
\begin{centering}
\includegraphics[width=3in,height=3in]{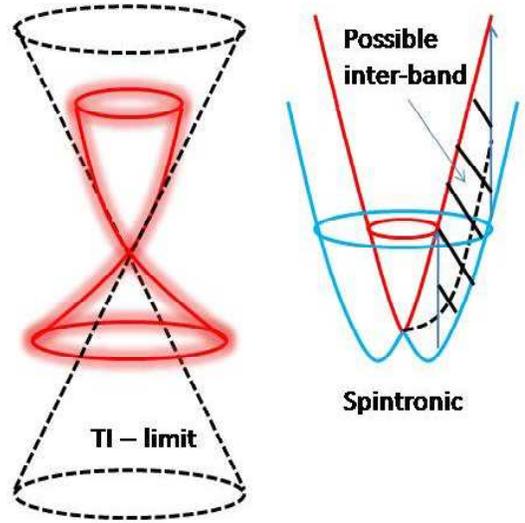} 
\par\end{centering}

\caption{(Color online) Schematics of dispersion curves used to describe a
topological insulator (left hand figure) and the opposite limit of
weak spin orbit coupling (right hand figure). }

\label{fig1} 
\end{figure}

The Hamiltonian on which this work is based takes on the form 
\begin{equation}
H_{0}=\frac{\hbar^{2}k^{2}}{2m}+\hbar v_{F}(k_{x}\sigma_{y}-k_{y}\sigma_{x})\label{H0}
\end{equation}
where $\mathbf{k}$ is momentum, $m$ is the Schr\"{o}dinger mass, $v_{F}$
the Dirac Fermi velocity and $\sigma_{x}$, $\sigma_{y}$ spin Pauli
matrices. For spintronic materials the first term in (1) is dominant
with the second small in comparison; while for topological insulator
it is the opposite. This is shown schematically in Fig.1. The dashed
black curve (left frame) is the perfect cone which would apply for
example to the Dirac fermions of graphene and is included here for
comparison with the heavy solid red curve which illustrates the dispersion
curve found in topological insulators. In this case the subdominant
Schr\"{o}dinger piece in the Hamiltonian (1) reshapes the cone into an
hourglass figure with upper conductance band (red) narrowing in cross-section
as compared with the Dirac cone and lower valence band (red) fanning
out from the cone giving it a larger cross-section. The right hand
frame illustrates the other limit of dominant Schr\"{o}dinger quadratic
band (black dashed curve) with subdominant spin orbit coupling. This
leads to a splitting of the dashed curve into two bands with heavy
solid red ($E_{+}$ ) contained inside the solid blue ($E_{-}$ )
dispersion curve.

When a magnetic field $B$ is applied perpendicular to the plane of
the two dimensional Hamiltonian (1), Laudau levels (LL) form and the
dynamic magneto-conductivity is given by \cite{Li2} 
\begin{eqnarray}
\sigma_{\alpha\beta}(\omega) & = & \frac{-i}{2\pi l_{B}^{2}}\sum_{N,N^{\prime},s,s^{\prime}}\frac{f_{N,s}-f_{N^{\prime},s^{\prime}}}{E_{N,s}-E_{N^{\prime},s^{\prime}}}\nonumber \\
 &  & \times\frac{\langle N,s|j_{\alpha}|N^{\prime},s^{\prime}\rangle\langle N^{\prime},s^{\prime}|j_{\beta}|N,s\rangle}{\omega-E_{N,s}+E_{N^{\prime},s^{\prime}}+i/(2\tau)}
\end{eqnarray}
where we have included a small phenomenological constant residual
broadening of $1/(2\tau)$. In Eq. (2) $l_{B}=1/\sqrt{e|B|/\hbar}$
is the magnetic length with $e$ the electron charge, $f$ the Fermi
distribution, $E_{N,s}$ the Landau level energies with eigenfunction
$|N,s\rangle$. The matrix elements $\langle N,s|j_{\alpha}|N^{\prime},s^{\prime}\rangle$
with $j_{\alpha}$ the $\alpha^{\prime}$ th component of the current
operator carry the information on the optical selection rules.
The Fermi function $f_{N,s}\equiv1/[e^{\beta(E_{N,s}-\mu)}+1]$ with
$\beta$ the inverse temperature and $\mu$ the chemical potential.
Details can be found in reference (26) with their Eq. (10) corrected
to include an overall additional minus sign in the second line. The
eigen energies are given by 
\begin{equation}
E_{N,s}=NE_{0}+s\sqrt{(E_{0}/2)^{2}+2NE_{1}^{2}}\label{landau}
\end{equation}
for $N=1,2,3...$, $s=\pm$ and 
\begin{equation}
E_{N=0}=E_{0}/2.
\end{equation}
Here $E_{1}=\hbar v_{F}\sqrt{e|B|/\hbar}$ is the Dirac magnetic energy
scale and $E_{0}=\hbar^{2}/(ml_{B}^{2})$ is the corresponding Schr\"{o}dinger
magnetic energy scale. We begin with the limit when, for a given value
of the magnetic field $B$, $E_{1}<E_{0}$. Reference (26) was exclusively
concerned with the opposite limit $E_{1}>E_{0}$ which applies to
topological insulators. When appropriate we provide comparisons between
these two cases. In a later section we also present additional new
results in the TI limit as well. When $E_{1}\rightarrow0$, the energies
$E_{N,s}\simeq(N+s/2)E_{0}+\frac{sE_{0}}{2}8NP$ with $P\equiv(E_{1}/E_{0})^{2}$
for $N=1,2,3...$ and $E_{N=0}=E_{0}/2$. Clearly in this case we
are dealing with two separate Schr\"{o}dinger Laudau level series with
the $N=0$ term included with $E_{N,+}$ ($N\geqslant1$). These two
series can be thought of as originating from the spin degeneracy of
a simple quadratic band split by a small spin-orbit coupling term
which shifts one band up and the other down in energy. This is illustrated
in Fig. 2 where we show the $E_{N,-}$ series as red dashed horizontal
lines and the $E_{0}$ plus $E_{N,+}$ as solid black horizintal lines.
\begin{figure}[tp]
\begin{centering}
\includegraphics[width=3in,height=3in]{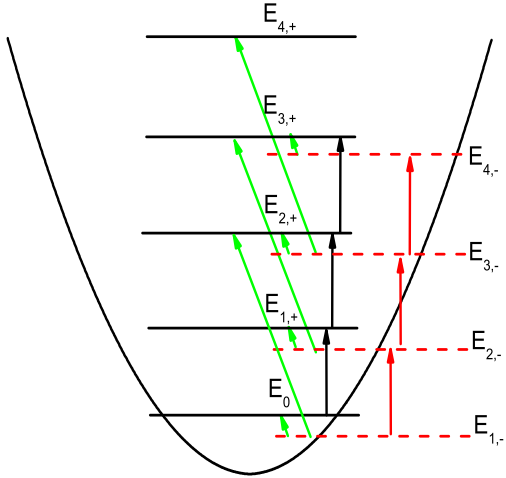} 
\par\end{centering}

\caption{(Color online) Schematic of possible optical transitions in the limit
of small spin orbit coupling. The intraband transitions ($++$ solid
black arrows) and ($--$ solid red arrows) have optical spectral weight
of order one and remain finite when $v_{F}=0$. The interband transitions
(solid green arrows) have weight of order $P^{2}$ for ($+-$) and
$P^{3}$ for ($-+$). }

\label{fig2} 
\end{figure}

The final form of the equations for the absorptive part of the dynamic
conductivity $Re\sigma_{xx}(\omega)$ and $Im\sigma_{xy}(\omega)$
respectively for longitudinal and Hall conductivity given in Eq. (22)
of reference (26) will be our starting point. They are 
\begin{eqnarray}
 &  & \mp\frac{e^{2}}{2\hbar}\sum_{s}(f_{0}-f_{1,s})\frac{F(0,s)}{H(0,s)}E_{0}\nonumber \\
 &  & \times[\delta(\hbar\omega-E_{0}H(0,s))\pm\delta(\hbar\omega+E_{0}H(0,s)]\nonumber \\
 &  & \mp\frac{e^{2}}{2\hbar}\sum_{N=1,s,s^{\prime}}(f_{N,s}-f_{N+1,s^{\prime}})\frac{F(N,s,s^{\prime})}{H(N,s,s^{\prime})}E_{0}\nonumber \\
 &  & \times[\delta(\hbar\omega-E_{0}H(N,s,s^{\prime}))\nonumber \\
 &  & \pm\delta(\hbar\omega+E_{0}H(N,s,s^{\prime})]\label{RXY}
\end{eqnarray}
where the upper sign applies to $Re\sigma_{xx}(\omega)$ and the lower
sign to $Im\sigma_{xy}(\omega)$. It is clear from the form of equation
(5) that the optical matrix elements have restricted the possible
transitions between LL to $N,s\rightarrow N+1,s^{\prime}$ with $F(N,s,s^{\prime})=\langle N,s|j_{\alpha}|N^{\prime},s^{\prime}\rangle\langle N^{\prime},s^{\prime}|j_{\beta}|N,s\rangle\pi$
and $H(N,s,s^{\prime})=-1+s\sqrt{1/4+2NP}-s^{\prime}\sqrt{1/4+2(N+1)P}$
related to the energies of the possible optical transitions. While
the complicated expressions for $F(N,s,s^{\prime})$ specified in
reference (26) (not repeated here as they are rather complicated and
not particularly illuminating) are to be used in the numerical results
that we will present later, it is helpful to start with approximate
expressions which apply in the limit of small $P$. Defining $\mathcal{F}(N,s,s^{\prime})\equiv F(N,s,s^{\prime})E_{0}/H(N,s,s^{\prime})$
we find for the intraband optical transitions to leading order ($N=1,2,...$)
and for $N=0$ \cite{Li1} 
\begin{equation}
\mathcal{F}(N,-,-)=-N/2(1-4NP^{2})E_{0}
\end{equation}
\begin{equation}
\mathcal{F}(N,+,+)=-(1+N)/2[1+4(N+1)P^{2}]E_{0}
\end{equation}
with next corrections higher order in $P$ and for the interband transition
\begin{equation}
\mathcal{F}(N,+,-)=2(1+2N)P^{2}E_{0}
\end{equation}
\begin{equation}
\mathcal{F}(N,-,+)=-8N(1+N)P^{3}E_{0}
\end{equation}

The first thing that needs to be emphasized is that the interband
transition carry little optical spectral weight in the limit $P\ll1$.
The ($+,-$) transitions go like the square of $P$ and correspond
to the absorption of photon that have small energies as seen in Fig.
2 (short green arrows). Such transition vanish as $P\rightarrow0$.
The ($-,+$) optical transitions (long green arrows) go like the cube
of $P$ and so are even less important. These correspond to finite
photon energies with limiting value of $2E_{0}$ as $P\rightarrow0$
as seen in Fig. 2. In contrast the intraband transitions remain finite
as $P\rightarrow0$ and so dominate the optical absorption. They represent
the only possible absorption processes in a pure non relativistic
system with no spin-orbit coupling. This is clear from Fig. 2 where
the black arrows indicate the ($+,+$) intraband and red arrows the
($-,-$) intraband transitions. For a typical spintronic material
at an applied magnetic field $B$ of one tesla, $E_{0}\simeq1.16$
meV for $m=0.1m_{e}$ (with $m_{e}$ the bare electron mass) and for
a Dirac Fermi velocity $v_{F}\simeq4.3\ast10^{3}$m/s, $E_{1}\simeq0.104$
meV which corresponds to a $P$ value of $\simeq0.008$. Even if the
magnetic field is reduced by a factor of fifty our expansion parameter
$P$ $\simeq0.4$ and consequently the interband optical transitions
are suppressed by a factor of $(0.4)^{2}=0.16$ which is still small.
It is important however to realize in this context that our expansion
in small $P$ implies that we are working at finite value of magnetic
field (here 1 Tesla for definiteness). Because $E_{1}$ goes like
$\sqrt{B}$ and $E_{0}$ goes instead like $B$, at very low field
the Dirac energy scale will always be larger than the Schr\"{o}dinger
magnetic energy scale and we are mapped out of the small $P$ regime.

\section{Numerical results}

\begin{figure}[tp]
\begin{centering}
\includegraphics[width=3.2in,height=3.2in]{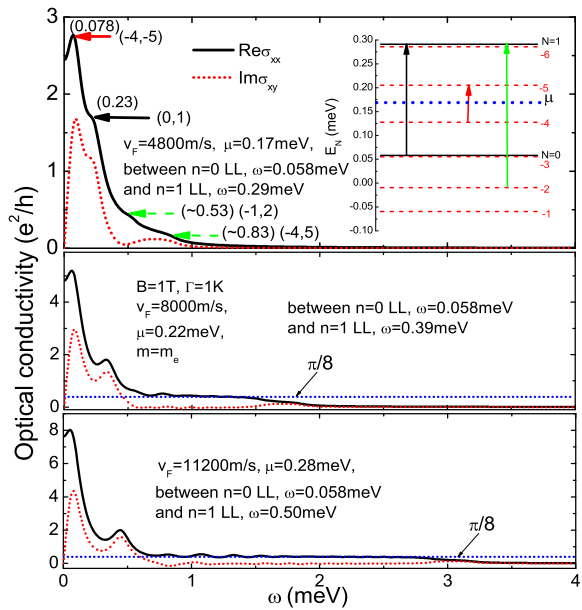} 
\par\end{centering}

\caption{(Color online) The dynamic longitudinal $\sigma_{xx}(\omega)$ (solid
black curve) and transverse $\sigma_{xy}(\omega)$ (Hall,dotted red
curve) conductivity in units of $e^{2}/h$ as a function of photon
energy $\hbar\omega$ in meV. In all cases the Schr\"{o}dinger mass $m$
was set at its free electron value $m_{e}$, the magnetic field at
$B=1$ Tesla, the broadening $\Gamma=1K$ and the chemical potential
$\mu$ is half way between the $N=0$ and $N=1$ landau level (LL).
The Schr\"{o}dinger magnetic energy scale $E_{0}=0.116$ meV while the
Dirac scale $E_{1}$ is varied and is 0.116 meV for $P=1$ in the top frame, 0.194 meV for P=2.8 in the middle frame and 0.272 meV for P=5.5 in the bottom frame. 
The corresponding Dirac Fermi velocities are $4.8\times10^{3}$
m/s, $8\times10^{3}$ m/s and $11.2\times10^{3}$ m/s respectively.
In the top frame arrows highlight the energies of some of the allowed
optical transitions which contribute to the dynamic conductivity.
The relevant LL energy scheme is given in the inset where the chemical
potential is shown as the horizontal dotted blue line. }

\label{fig3} 
\end{figure}

In Fig. 3 we show results of complete numerical calculations based
on Eq. (5) without making the approximation that $P$ be very small
and using the exact expressions for the optical matrix elements in
$F(N,s,s^{\prime})$.\cite{Li2} In the top frame we show $Re\sigma_{xx}(\omega)$
(solid black curve) and $Im\sigma_{xy}(\omega)$ (dotted red curve)
in units of $e^{2}/h$ as a function of photon energy $\omega$ in
meV. Here $E_{0}=E_{1}=0.116$meV, $v_{F}=4.8\times10^{3}$ m/s with
chemical potential $\mu$ set half way between the $N=0$ and $N=1$
LL ($\mu=0.17$ meV). We have also used a residual scattering $\Gamma=1/2\tau=1K$
which broadens out the LL although structures corresponding to the
allowed optical transitions between levels are still seen and some
of these are identified by arrows. Black and red are intraband, green
interband in both main frame and on energy level inset. The inset
gives the details of the allowed optical transitions including the
position of the chemical potential at $.17$ meV (dotted blue horizontal
line). We first note that $E_{N-}$ is negative only for $N=1$ and
$2$ and is near the level with $N=0$ for $N=3$. It becomes
larger than the value of the chemical potential for $N=5$ which means
that this level is unoccupied and hence the interband transition are
limited to $7$ with the $-4\rightarrow5$ corresponding to a photon
energy of $.82$ meV which is seen as a cut off for both longitudinal
and transverse optical conductivity as indicated by a green arrow.
The lowest energy transition is intraband at $0.078$ meV and the
second intraband is at $0.23$ shown by the red and black arrows in
the inset. Only $9$ transitions are possible, two intraband and seven
interband, one shown as the green arrow in the level scheme diagram.
This is in sharp contrast to the pure Dirac case for which the interband
transitions have no natural cut off in our continuum limit Hamiltonian
(1). Note that the magnetic energy scale is small and that, even for
$\Gamma=1K$ which is certainly reasonable, the broadening has eliminated
the sharp peaks at the LL energies. Finally we note that, by choice,
$P$ was set to be 1 so that we are far away from the $P\ll1$ limit
considered in the simplified Eq. (6) to (9). Nevertheless, the interband
transition still correspond to suppressed optical spectral weight.
In the second frame of Fig. 3 we have increased the value of $P$
to 2.8 and $\mu=0.22$ meV which is again half way between $N=0$
and $N=1$ Landau level. Now we see the emerging of a prominent and
distinct interband background in $Re\sigma_{xx}(\omega)$ on which
is superimposed very broaden LL peaks. This background extends to almost
$\lesssim2.0$ meV and its height in units of $e^{2}/h$ is exactly
$\pi/8$ (shown as the dotted blue line). Remarkably this is precisely the value found in graphene
except for the factor of 4 which accounts for valley and spin degeneracy.
The interband background is even more prominently developed in the
lower frame of Fig. 3 for $P=5.5$. Here the cutoff is $\lesssim3.0$
meV.

We can gain insight into these results by comparing with results obtained
in the topological insulator limit as we will do in the next section.
Before we do this however it is also useful to consider the limit
of zero magnetic field. In that case the formula for $Re\sigma_{xx}(\omega)$
is given by 
\begin{eqnarray}
 &  & \frac{e^{2}}{\omega}\frac{1}{4\pi^{2}}\int_{0}^{\infty}[f(E_{-}(k))-f(E_{+}(k))]kdk\notag\nonumber \\
 &  & (\hbar v_{F})^{2}\int_{0}^{2\pi}d\theta\pi\delta(\hbar\omega-E_{+}(k)+E_{-}(k))
\end{eqnarray}
with 
\begin{equation}
E_{s}(k)=\frac{\hbar^{2}k^{2}}{2m}+s\hbar v_{F}k
\end{equation}
\begin{figure}[tp]
\begin{centering}
\includegraphics[width=3in,height=3in]{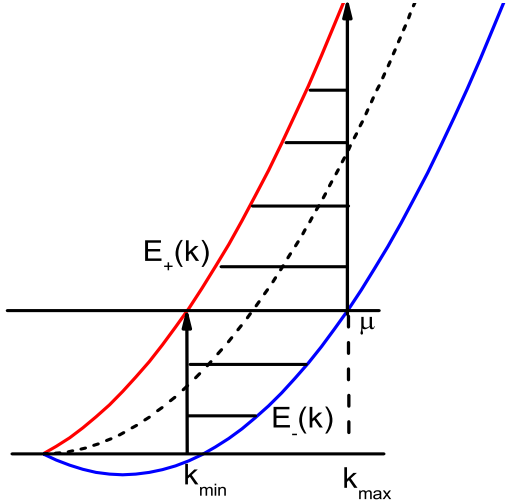} 
\par\end{centering}
\caption{(Color online) The pure Schr\"{o}dinger quadratic in $k$ dispersion curve
(dashed line) compared with $E_{+}(k)$ (solid red) and $E_{-}(k)$
(solid blue) dispersion. The vertical arrows show the two extremum
interband transitions that are possible from which we identify $k_{min}$
and $k_{max}$. The shaded region emphasizes the area associated with
such allowed transitions. }
\label{fig4} 
\end{figure}
The first thing to note about Eq. (10) is that the Schr\"{o}dinger contribution
to the energy drops out of the interband energy difference $[E_{+}(k)-E_{-}(k)]$
but it does remain in the thermal factor $[f(E_{-}(k))-f(E_{+}(k))]$.
At zero temperature this factor reduces either to one in the interval
for which interband transitions are possible or is otherwise zero.
The kinematics involved are shown in Fig. 4. The red solid line gives
$E_{+}(k)$ vs $k$, the blue is $E_{-}(k)$ and the black dashed curve,
which is for comparison, gives the dispersion curve when the spin-orbit
term is zero i.e. $v_{F}=0$ in Eq. (11). The shaded region shows
the possible interband transitions and defines a minimum and a maximum
value of momentum $k_{\min}$ and $k_{\max}$, which provide lower
and upper limits on the integral in Eq. (10). These cut offs are due
to the thermal factors which require an occupied initial state and
an unoccupied final state. For graphene $k_{\min}$ corresponds to
the onset of interband absorption at a photon energy of exactly twice
chemical potential $2\mu$. In that case however $k_{\max}$ does
not exist, since $E_{-}(k)$ is negative for all values of $k$ and
never crosses the chemical potential ($\mu>0$). Applying these restriction
to Eq. (11) means that interband optical transitions are only possible
for energies between 
\begin{equation}
\omega_{\min}=E_{+}(k_{\min})-E_{-}(k_{\min})
\end{equation}
and 
\begin{equation}
\omega_{\max}=E_{+}(k_{\max})-E_{-}(k_{\max}).
\end{equation}
But the application of these cut off on momentum in Eq. (10) is the
only effect of the thermal factors which, at zero temperature, have
magnitude of one. This means that the remaining integral is identical
to that for graphene leading to precisely the same absolute value
of the universal background. Of course now the background is restricted
to the interval $\omega_{\min}$ to $\omega_{\max}$. Thus $Re\sigma_{xx}(\omega)$
for interband absorption is still equal to $e^{2}\pi/8h$ but is non
zero only for photon energies $\omega_{\min}\leqslant\omega\leqslant\omega_{\max}$.
This remarkable result is consistent with the general trends found
in our numerical work presented in Fig. 3 for the case of a finite
$B$. Simple formulas for $\omega_{\min}$ and $\omega_{\max}$ are
easily obtained 
\begin{equation}
\omega_{\min}=-2mv_{F}^{2}+2\hbar v_{F}\sqrt{(\frac{mv_{F}}{\hbar})^{2}+\frac{2m\mu}{\hbar^{2}}}
\end{equation}
\begin{equation}
\omega_{\max}=2mv_{F}^{2}+2\hbar v_{F}\sqrt{(\frac{mv_{F}}{\hbar})^{2}+\frac{2m\mu}{\hbar^{2}}}
\end{equation}
We can check that in the limit $m\rightarrow\infty$ (pure relativistic
case) $\omega_{\max}=\infty$ and $\omega_{\min}=2\mu$. In the opposite
limit of $v_{F}\rightarrow0$ both $\omega_{\min}$ and $\omega_{\max}$
are zero and there are no interband transitions. In all cases the
width of the photon window over which the universal background has
height $e^{2}\pi/8h$ is $\omega_{\max}-\omega_{\min}=4mv_{F}^{2}$
which is linear in $m$ and quadratic in $v_{F}$, and is independent
of the chemical potential $\mu$. However, both the onset and termination
of the interband background do depend on the chemical potential $\mu$
which appears in the square root factor of Eq. (14) and this shifts
both upper and lower limits on the interband absorption. A similar
effect holds when a magnetic field is additionally applied, as we
will describe below. First we note that the total optical spectral
weight ($W_{IB}$) in the interband background is $W_{IB}=(e^{2}\pi/8h\times4mv_{F}^{2})$.
In addition there are the intraband transitions which give the Drude
response. In the appendix we show that the optical spectral weight
contained in the Drude when $B=0$ is given by 
\begin{eqnarray}
 &  & \frac{e^{2}m(\sqrt{\hbar^{2}v_{F}^{2}+(2\hbar^{2}/m)\mu}-\hbar v_{F})\sqrt{\hbar^{2}v_{F}^{2}+(2\hbar^{2}/m)\mu}}{8\hbar^{3}}+\notag\nonumber \\
 &  & \frac{e^{2}m(\sqrt{\hbar^{2}v_{F}^{2}+(2\hbar^{2}/m)\mu}+\hbar v_{F})\sqrt{\hbar^{2}v_{F}^{2}+(2\hbar^{2}/m)\mu}}{8\hbar^{3}}
\end{eqnarray}
where the first line comes from the ($+,+$) transitions and the second
from the ($-,-$) transitions. Formula (16) applies also to the opposite
limit of a topological insulator but in this case only the first line
is retained since only the ($+,+$) transitions are possible. For
the non-relativistic limit the Drude weight $W_{D}$ is 
\begin{equation}
W_{D}=\frac{e^{2}\pi}{h}[\mu+\frac{1}{2}mv_{F}^{2}].
\end{equation}
Thus $W_{D}$ does depend on the chemical potential $\mu$ in contrast
to the interband contribution $W_{IB}$. The total optical spectral
weight $W_{T}=W_{D}+W_{IB}=\frac{e^{2}\pi}{h}[\mu+mv_{F}^{2}].$ We
will return to this fact later when we consider spectral weight redistribution
between intra- and inter-band transition as $\mu$ is varied and compare
with the case of topological insulators.
\begin{figure}[tp]
\begin{centering}
\includegraphics[width=3.5in,height=3.3in]{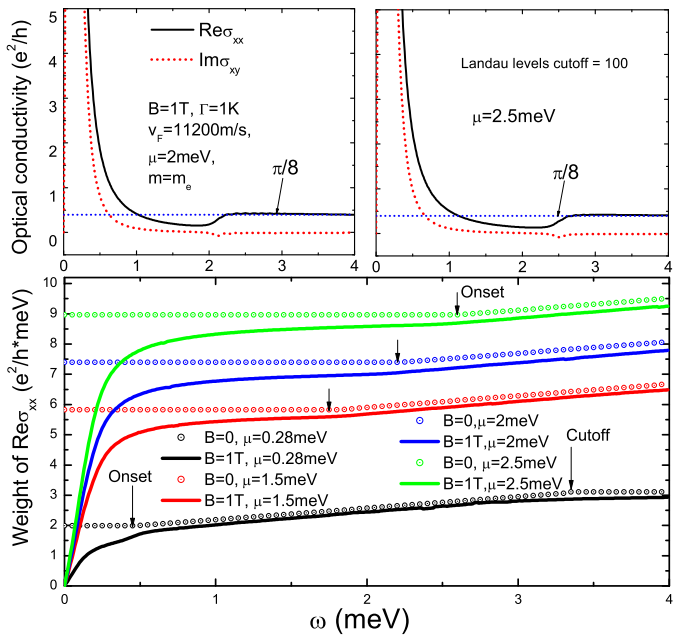} 
\par\end{centering}
\caption{(Color online) Top frame, the dynamic longitudinal $\sigma_{xx}(\omega)$
(solid black curve) and transverse $\sigma_{xy}(\omega)$ (Hall, dotted
red curve) conductivity in units of $e^{2}/h$ as a function of photon
energy $\hbar\omega$ in meV. Parameters are the same as for the bottom frame of Fig.3
except that different values of chemical potential are employed namely
$\mu=2.0$ meV and $\mu=2.5$ meV. Bottom frame, the optical sum $W(\omega_{c})$
as a function of upper cut off $\omega_{c}$ of the longitudinal conductivity
$\sigma_{xx}(\omega)$. The four curves are for $\mu=0.28$ meV (black),
$\mu=1.5$ meV (red), $\mu=2.0$ meV (blue) and $\mu=2.5$ meV (green).
The dotted curves are for the case of $B=0$ and are included for
comparison with the solid curves for $B=1$ Tesla. }
\label{fig5} 
\end{figure}
In the top frame of Fig. 5 we show results for the real part of the
longitudinal dynamic conductivity ($Re\sigma_{xx}(\omega)$, solid
black curve) and the imaginary part of the transverse Hall conductivity
($Im\sigma_{xy}(\omega)$, dotted red curve) for the same parameters
as in the bottom frame of Fig. 3 but now the chemical potential has
been increased. In the left frame $\mu=2.0$ meV and in the right
frame $\mu=2.5$ meV. Increasing $\mu$ eliminates several interband
transitions which now are no longer possible because of Pauli blocking.
The optical spectral weight of these interband transitions has been
transferred partly to the low energies intraband peak which becomes
more prominent with increasing $\mu$. In the lower frame of Fig.
3 a background of height $\frac{e^{2}\pi}{8h}$ is well defined in
the longitudinal conductivity down to photon energies $\lesssim1$
meV. On the other hand in the top frame of Fig. 5 this background
is very suppressed in the region 1 meV to $\thicksim\mu$ where it
is seen to jump to its universal value. For the case of $B=0$ we
saw that this threshold energy is $\omega_{\min}$, defined in Eq.
(14). This universal background then remains up to an upper cut off
$\omega_{\max}$ defined in Eq. (15) for $B=0$ at which point it
drops to zero. For finite $B$ these critical energies are some what
different from those of Eq. (14) and vary with the magnitude of $B$.
In Fig. 5 (top frame) $B=1$ Tesla. Increasing $B$ increases the
distance in energy between the various Landau levels and so the energy
of the minimum and maximum possible interband transition is changed.

A more quantitative look at spectral weight distribution between inter
and intra band optical transition and its variation with value of
the chemical potential $\mu$ is given in the lower frame of Fig.
5. There we present results for the total spectral weight contained
in the longitudinal conductivity below a variable upper photon energy
$\omega$ defined as 
\begin{equation}
W(\omega)=\int_{0}^{\omega}Re\sigma_{xx}(\omega^{\prime})d\omega^{\prime}
\end{equation}
The units on $W(\omega)$ are $\frac{e^{2}}{h}$ times meV's and $\omega$
on the horizontal axis is in meV. Four values of chemical potential
$\mu$ are shown $\mu=0.28$ meV (black), $\mu=1.5$ meV (red) $\mu=2.0$
meV (blue) and $\mu=2.5$ meV (green). Here the residual scattering
rate $\Gamma=1K$, the Fermi velocity $v_{F}=11200$ m/s and the Schr\"{o}dinger
mass $m$ is equal to the bare electron mass ($m_{e}$).

The solid curves are for a magnetic field of $B=1$ Tesla while the
color coded open circles which are shown for comparison, are for zero
magnetic field. For the first black curve with small value of chemical
potential $\mu$, inter and intra band contributions to $Re\sigma_{xx}(\omega)$
and to its integrated spectral weight $W(\omega)$ are not easily
distinguished although there is a clear onset, slightly below $\omega$=0.5
meV, indicated by the black arrow, and there is also a cut off energy
around 3.4 meV (see black arrow) beyond which the integrated spectral
weight $W(\omega)$ ceases to increase. This is seen particularly
clearly in the curve for $B=0$ . For the other three cases considered,
the intra and interband transitions are well separated. Below the
onset of the interband transition at $\omega_{min}$ of Eq. (14),
the $B=0$ curve is completely flat and contains the Drude spectral
weight $W_{D}$ of formula (17). At $\omega=\omega_{min}$ there is
an abrupt change in slope from zero to a finite amount set by the
value of the universal background $\frac{e^{2}\pi}{8h}$. This kink,
which we marked with a black arrow, while most pronounced in our $B=0$
curves is also present at a slightly different energy in our $B=1$
Tesla curves. This is expected and is traced to our numerical results
for the absorptive part of the longitudinal conductivity shown in
the top frame of Fig. 5 where the onset of the universal interband
background is clearly seen around $\omega=\omega_{min}$ . The intraband
transitions below this onset provide a peak in $Re\sigma_{xx}(\omega)$
, displayed from $\omega=0$, which has however decayed to very small
values before $\omega=\omega_{min}$ is reached. It is the displacement
of the intraband transition (which provides a Drude when B=0) to finite
frequency in a nonzero magnetic field (B=1) which accounts for the
gradual increase in $W(\omega)$ out of zero energy ($\omega=0$)
towards the Drude plateau of the B=0 case which is now never perfectly
flat.

Beyond the region of linear increase in $W(\omega)$, in our B=0 case
the optical spectral weight reaches saturation as can be seen in the
black curves for which $\omega_{\max}$ of Eq. (15) falls within the
range of energies shown in the figure. For the other cases one needs
to go to energies beyond 4 meV. The saturated value for B=0 is the
sum of the Drude contribution of Eq. (17) plus the amount in the interband
which add up to $\frac{e^{2}\pi}{h}[\mu+mv_{F}^{2}]$.

This behavior is very different from what is known for the pure relativistic
case such as for graphene for which no upper cutoff exists other than
the Brillouin zone cut off. While we will see in a later section that
modifications of the known graphene behavior also arise when a small
subdominant Schr\"{o}dinger piece is added to the dominant Dirac contribution,
for the pure relativistic case spectral weight is simply redistributed
between inter and intra contributions as $\mu$ is varied. Here we
see that for materials with both Dirac and Schr\"{o}dinger pieces in the
Hamiltonian there is no strict conservation of spectral weight with
changes in $\mu$, an issue we will return to in section V. While
only $\mu$ changes in the various curves shown in the lower frame
of Fig. 5, they do not merge as $\omega$ gets large.

\section{Comparison with a Topological Insulator}

We next compare the results of Fig. 3 with results obtained in the
topological insulator limit shown in Fig. 6. The three top frames
give the absorptive part of the diagonal conductivity $Re\sigma_{xx}(\omega)$
(solid black curve) and comparison with the imaginary part of transverse
conductivity $Im\sigma_{xy}(\omega)$ (dotted red curve) in units
of $e^{2}/h$ as a function of photon energy $\omega$ in meV. We
have taken the Fermi velocity to be $4.3\times10^{5}$ m/s, the magnetic
field to be 1 Tesla, the scattering rate $\Gamma=15K$ and the Schr\"{o}dinger
mass equal to the bare electron mass ($m_{e}$) in the third lowest
frame, $m=0.1m_{e}$ in the second lowest frame and $m=0.05m_{e}$
in the top frame. Decreasing $m$ moves us further away from the pure
Dirac case which would correspond to graphene. In all these frames
the chemical potential was set to fall halfway between $N=1$ and
$N=2$ Landau level. The first thing to note about these results is
that the magnetic energy scale associated with the Landau levels is
much larger than that in Fig. 3 for the non-relativistic limit. Here
the relativistic energy scale given by the Dirac term only is $E_{1}=10.4meV$
while the non relativity magnetic energy scale for $m=m_{e}$ is 0.116
meV, two orders of magnitude smaller. Secondly the interband optical
peaks in $Re\sigma_{xx}(\omega)$ remain to very high energies as
in the case of graphene previously discussed in Ref. (17). Just as
we found it enlightening in Fig. 3 to consider several values of $v_{F}$
for fixed value of $m$, here we fix $v_{F}$ and consider 3 values
of $m$. For $m=m_{e}$ third frame from top in Fig. 6 the results
are not significantly different from those in pure graphene.\ As
$m$ is decreased to $0.1m_{e}$ a splitting of the peaks in $Re\sigma_{xx}(\omega)$
into pairs is seen. This arises because the Schr\"{o}dinger term breaks
the particle hole symmetry of the pure relativistic case and the optical
transition $-N$ to $N+1$ no longer has the same energy as $-(N+1)$
to $N$. This splitting is even more pronounced when $m$ is decreased
further to $m=0.05m_{e}$ as in the top frame. In this case not all
peaks can be easily identified as split pairs. Of course even for
$m=m_{e}$, (third frame down) there should in principle be a splitting
but here we have taken a smearing parameter $\Gamma=15K$ which is
enough to merge them so that a single peak is effectively seen.

We will return to the results of Fig. 6 later. First it is useful
to consider analytic results which we can obtain only in the limit
of $P^{-1}\ll1$ when the non relativistic correction to pure Dirac
is small. In this case the optical spectral weight in units of $e^{2}/2\hbar$
which is associated with the allowed transitions at zero temperature
have the approximate form\cite{Li1,Li2}

\begin{eqnarray}
 &  & \mathcal{F}(N,-,-)=[\frac{\sqrt{N}+\sqrt{N+1}}{4\sqrt{2}}\nonumber \\
 &  & -\frac{1/\sqrt{P}}{16[\sqrt{N(1+N)}-2\frac{N(N+1)}{1+2N}]}]E_{1}
\end{eqnarray}
which will not be needed for $\mu\geqslant0$ because such transitions
are Pauli blocked 
\begin{eqnarray}
 & \mathcal{} & \mathcal{F}(N,+,+)=-[\frac{1}{4\sqrt{2}[-\sqrt{N}+\sqrt{N+1}]}\notag\nonumber \\
 &  & +\frac{1/\sqrt{P}}{16[\sqrt{N(1+N)}-2\frac{N(N+1)}{1+2N}]}]E_{1}
\end{eqnarray}
\begin{eqnarray}
 & \mathcal{} & \mathcal{F}(N,-,+)=[\frac{-1}{4\sqrt{2}[\sqrt{N}+\sqrt{N+1}]}\notag\nonumber \\
 &  & +\frac{1/\sqrt{P}}{16[\sqrt{N(1+N)}+2\frac{N(N+1)}{1+2N}]}]E_{1}
\end{eqnarray}
and 
\begin{eqnarray}
 & \mathcal{} & \mathcal{F}(N,+,-)=[\frac{1}{4\sqrt{2}[\sqrt{N}+\sqrt{N+1}]}\notag\nonumber \\
 &  & +\frac{1/\sqrt{P}}{16[\sqrt{N(1+N)}+2\frac{N(N+1)}{1+2N}]}]E_{1}
\end{eqnarray}
where we are working to lowest order in $1/\sqrt{P}$. This provides
the first correction to the pure relativistic limit. Note that, in addition to the $\mathcal{F}(N,s,s')$, 
the absorptive part of the longitudinal optical conductivity $Re\sigma_{xx}(\omega')$ depends additionally on thermal factors which give a $+1$ or $-1$ at$T=0$.
consequently the optical spectral weight under an allowed optical transition between LL which must be positive, is in all cases equal to the absolute value of 
$\mathcal{F}(N,s,s')$. This means that the spectral weight corresponding to (20) and (21) carries an additional minus sign so that the optical spectral weight 
associated with (21) is reduced over its pure Dirac limit value while that associated with (22) is increased by exactly the same amount. When this correction
is dropped, we recover the result of Gusynin, Sharapov and Carbotte\cite{Carbotte1}
for graphene except for a missing factor of four accounting for spin
and valley degeneracy not present for topological insulators. Taking
the limit of large $N$ the optical spectral weight contained in the
intraband transitions is given by$\frac{e^{2}}{2\hbar}\mathcal{F}(N,+,+)_{N\rightarrow\text{large}}$.
Noting that $\frac{1}{[-\sqrt{N}+\sqrt{N+1}]}\simeq2\sqrt{N}$ and
that $\frac{1}{[\sqrt{N(1+N)}-2\frac{N(N+1)}{1+2N}]}\simeq2N$ we
get 
\begin{equation}
-\frac{e^{2}}{2\hbar}\mathcal{F}(N,+,+)_{N\rightarrow\text{large}}=[\frac{\sqrt{2N}}{4}+1/\sqrt{P}\frac{N}{2}]E_{1}
\end{equation}
But $\mu\simeq\sqrt{2N}E_{1}+E_{0}(\frac{\mu}{E_{1}\sqrt{2}})^{2}$
or $\sqrt{2N}\simeq\frac{\mu}{E_{1}}(1-\frac{\mu}{2mv_{F}^{2}})$
so that 
\begin{equation}
-\frac{e^{2}}{2\hbar}\mathcal{F}(N,+,+)_{N\rightarrow\text{large}}\simeq\frac{\mu}{4}(1+\frac{\mu}{2mv_{F}^{2}})\frac{e^{2}\pi}{h}
\end{equation}
In the limit of $m=\infty$ this reduces to $\frac{e^{2}\pi}{4h}\mu$
which is exactly the amount of optical spectral weight there would
be under the Drude in the $B=0$ limit in graphene except for the
factor of 4 in the denominator which would be canceled by a degeneracy
factor for two spins and two valleys. When $m$ is large but not infinite
the spectral weight under the intraband transition line is increased
from a normalized value of 1 by an amount $\frac{\mu}{2mv_{F}^{2}}$.
For our expansion to be valid we still need $\frac{\mu}{4mv_{F}^{2}}$
to be small which is a more restrictive condition than simply $1/\sqrt{P}\ll1$.
Nevertheless it shows clearly that the optical spectral weight residing
in the various optical lines is changed from the pure Dirac case when
a subdominant Schr\"{o}dinger piece is also present in the Hamiltonian.

In the large $N$ limit the intraband line is the cyclotron resonance
line of semiclassical theory. The cyclotron frequency was worked out
in Ref. {[}26{]} and found to be to lowest order correction for a
small Schr\"{o}dinger contribution,

\begin{equation}
\hbar\omega_{c}=\frac{E_{1}^{2}}{\mu}[1+\frac{3}{2}\mu/(mv_{F}^{2})]
\end{equation}
So that $\hbar\omega_{c}$ is increased for $m\neq\infty$ as is the
spectral weight under this line. It is interesting to compare the
spectral weight of the cyclotron resonance line with the Drude weight
($W_{D}$) for the zero magnetic field case. The expression for $W_{D}$
is (Eq. (16), first line only) 
\begin{equation}
\frac{e^{2}m[\hbar^{2}v_{F}^{2}+(2\hbar^{2}/m)\mu-\hbar v_{F}\sqrt{\hbar^{2}v_{F}^{2}+(2\hbar^{2}/m)\mu}]}{8\hbar^{3}}
\end{equation}
which is valid for any value of $m$. Assuming $m$ to be large but
not infinite we can expand (26) and obtain a first correction to pure
Dirac, we get

\begin{equation}
W_{D}=\frac{e^{2}\pi}{4h}\mu\lbrack1+\frac{\mu}{2mv_{F}^{2}}]
\end{equation}
which agrees perfectly with the spectral weight under the semiclassical
cyclotron line in this approximation.

In the comparisons made so far between the pure Dirac case and a topological
insulator with finite Schr\"{o}dinger contribution we have considered
the chemical potential as fixed. It is important to realize that this
does not correspond to a fix doping as $m$ is varied. For a fix density
of charge carriers(n) away from the neutrality point we have that 
\begin{equation}
n=\int_{0}^{k_{\min}}\frac{kdk}{2\pi}=\frac{1}{4\pi}[\frac{-mv_{F}}{\hbar}+\frac{mv_{F}}{\hbar}\sqrt{1+\frac{2\mu}{mv_{F}^{2}}}]^{2}
\end{equation}
where we have used $k_{\min}$ determined from Fig. 4. Eq. (28) holds
whatever may be the value of $m$. For large $m$ retaining the first
leading correction we get 
\begin{equation}
n\simeq\frac{1}{4\pi}(\frac{\mu}{\hbar v_{F}})^{2}(1-\frac{\mu}{mv_{F}^{2}})
\end{equation}

from which it follows that 
\begin{equation}
\mu\simeq\sqrt{4\pi n}\hbar v_{F}(1+\frac{\sqrt{\pi n}\hbar}{mv_{F}})
\end{equation}
For $m=\infty$ we recover the known result for pure Dirac fermions
$\mu=\sqrt{4\pi n}\hbar v_{F}$. For $m$ not infinite, $\mu$ is increased
over the pure relativistic limit value because of the change in the
electronic dispersion curves. This increase in chemical potential
is a direct consequence of the narrowing of the cross-section of the
conduction band cone due to the subdominant Schr\"{o}dinger term.

Returning to our approximate equations for $\mathcal{F}(N,+,+)$,
$\mathcal{F}(N,+,-)$ and $\mathcal{F}(N,-,+)$ we note that in all
cases the corrections for a non zero value of $1/P$ is of order $1/\sqrt{P}$
and more importantly, that the spectral weight associated with the
interband transition remains finite even when $1/\sqrt{P}=0$. This
is in sharp contrast to what we found in the non-relativistic limit
where they vanish when no subdominant Dirac correction is included.
For the pure relativistic case the photon energy associated with both
($+,-$) and ($-,+$) interband transitions is the same. But here
with $1/\sqrt{P}\neq0$ the ($+,-$) transition is slightly shorter
than is the ($-,+$) transition. We also see, noting the sign of the thermal factors in Eq. (5), that the optical spectral
weight associated with the larger photon energy has the smallest spectral
weight. To first leading order in $1/\sqrt{P}$ however the total
spectral weight under the two split lines is unchanged from the pure
relativistic case where $1/\sqrt{P}=0$. In graphene it is well known\cite{Carbotte1}
that as $\mu$ is increased so as to cross the $N$'th LL line only
one half the spectral weight remains in the $N$'th line and all others
for $n<N$ have disappeared with the entire spectral weight lost in
the interband transition reappearing in the single intraband line
which moves to lower photon energy with increasing $\mu$ and picks
up more intensity. This sum rule on the redistribution of the spectral
weight between inter and intra lines is encapsuled in the equation\cite{Carbotte1}
\begin{equation}
\sum_{n=0}^{N-1}\frac{2}{\sqrt{n+1}+\sqrt{n}}+\frac{1}{\sqrt{N+1}+\sqrt{N}}=\frac{1}{\sqrt{N+1}-\sqrt{N}}
\end{equation}
\begin{figure}[tp]
\begin{centering}
\includegraphics[width=3in,height=4.2in]{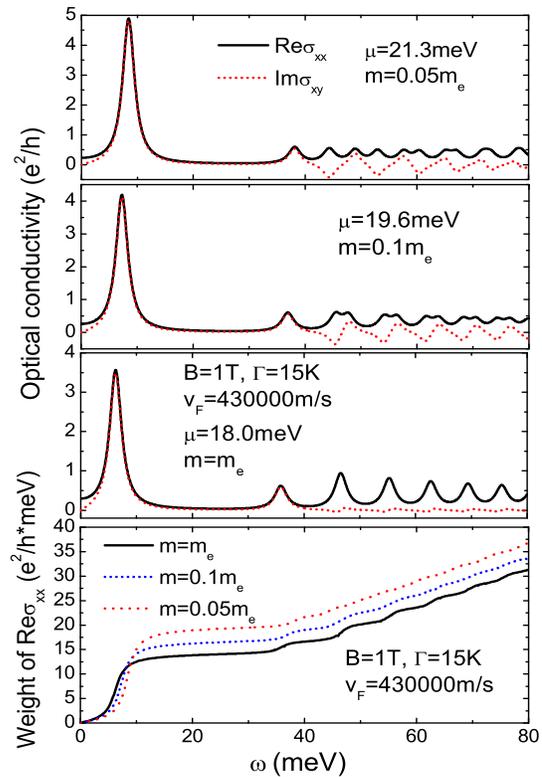} 
\par\end{centering}
\caption{(Color online) Top frame is the dynamic longitudinal $\sigma_{xx}(\omega)$
(solid black) and transverse $\sigma_{xy}(\omega)$ (Hall, dotted
red) conductivity in units of $e^{2}/h$ as a function of photon energy
$\hbar\omega$ in meV. The Dirac fermi velocity is set at $v_{F}=4.3\times10^{5}$
m/s which is nearly two orders of magnitude larger than used in the
top frame of Fig. 3 and is representative of topological insulators.
Top, second and third frames are for the Schr\"{o}dinger mass equal to
$m_{e}$, $0.1m_{e}$ and $0.05m_{e}$ respectively with $m_{e}$
the bare electron mass. The bottom frame gives the optical spectral
weight $W(\omega_{c})$ in units of $e^{2}/h\times$meV as a function
of photon energy $\hbar\omega$ in meV. The curves are for $B=1$
Tesla. The solid black is for $m=m_{e}$, dotted blue for $m=0.1m_{e}$
and dotted red is for $m=0.05m_{e}$.}
\label{fig6} 
\end{figure}
derived in Ref [17], see their Eq. (33). The first term on the
left is related to the spectral weight associated with all lines that
have completely disappeared while the second is half of the spectral
weight of the line $n=N$. The right hand side is the spectral weight
in the intraband line. It is clear from our approximate Eq. (19-22) and Eq. (5)
that this sum rule will no longer hold when $1/\sqrt{P}\neq0$ since
pairs of lines that have completely disappeared will have the same
spectral weight as for the pure Dirac case while the one for $n=N$
will have a correction of order $1/\sqrt{P}$ (reduction) which does not match
precisely the intraband correction of order $1/\sqrt{P}$. 

Next we return to our numerical results of Fig. 6 which do not require the
$1/\sqrt{P}\rightarrow0$ limit and these show explicitly a change in optical sum with Schr\"{o}dinger admixture. The first peak in each of the three top curves for $Re\sigma_{xx}(\omega)$
come from the intraband optical transitions while all other peaks
are interband. As the Schr\"{o}dinger contribution is increased through
a decrease in $m$ all intraband lines are seen to shift slightly to higher energies
as compared to the pure Dirac case (which close to the $m=m_{e}$
curves). Also the optical spectral weight contained
in the intraband line increases as the Schr\"{o}dinger mass is decreased.
This fact is more easily seen in the lower frame of Fig. 6 where we
give results for $W(\omega)$ of Eq. (18) as a function of $\omega$
in meV with $W$ in units of $\frac{e^{2}}{h}$meV. All the curves
start from zero and rise rapidly as we integrate over the intraband
line after which it has a plateau which is nearly but not quite constant,
followed by a more rapid rise modulated by small wiggles that reflect
the LL structure. These wiggles would of course be more pronounced
if we had reduced the broadening $\Gamma$. Here it is $15K$. The height of the
first plateau in $W(\omega)$ gives the amount of spectral weight
contained in the intraband transition and this clearly increase as
we move further away from the pure relativistic case. Differences
in $W(\omega)$ remain to high energies and these reflect the small
Schr\"{o}dinger admixture which we have added to the Hamiltonian.

\section{Redistribution of spectral weight with $\mu$}

Next we return to the issue of how optical spectral weight gets redistributed between
inter and intra band when the chemical potential $\mu$ is varied
i.e. a change in doping, without assuming $1/\sqrt{P}<<1$. We begin with the case of $B=0$ because
in that instance, as we have seen, we can get analytic results. In
previous sections we showed that the Drude weight is given by Eq.
(16). Only the first line in this equation applies in the relativistic
limit while both lines (Eq. (16)) contribute in the non-relativistic
case. Further, we saw that the interband transition provide a constant
background of $\frac{e^{2}\pi}{8h}$ in magnitude, with sharp absorption
edge at $\omega_{\min}$ which terminates at $\omega_{\max}$ (see
Eq. (14) and (15)). The total optical spectral weight in this background
is $4mv_{F}^{2}\frac{e^{2}\pi}{8h}$. For comparison with the relativistic
limit let us begin by computing the optical spectral weight that is
missing from the universal background between $\omega=0$ and $\omega=\omega_{\min}$.
For $m\rightarrow\infty$ we can expand the expression for $\omega_{\min}$
and obtain to lowest order 
\begin{equation}
\omega_{\max}\simeq2\mu\lbrack1-\frac{\mu}{2mv_{F}^{2}}]
\end{equation}
which gives the known result that the interband absorption edge starts
precisely at $2\mu$ in the pure relativistic limit. When $m$ is
not infinite the absorption edge has shifted to an energy somewhat
less than $2\mu$. The missing optical spectral weight is 
\begin{equation}
\frac{e^{2}\pi}{8h}\times2\mu\lbrack1-\frac{\mu}{2mv_{F}^{2}}]
\end{equation}
which is also somewhat less than $\frac{e^{2}\pi}{4h}\mu$ found in
the limit $m=\infty$. Now in the limit of a topological insulator
the Drude weight is given to leading
order in an expansion for large $m$ in Eq. (27) and is not equal to the missing weight in the universal background
given in Eq. (33). So the spectral weight redistribution sum rule
which is operative in the pure relativistic limit breaks down when
a subdominant Schr\"{o}dinger piece is added to the Hamiltonian.

The opposite limit (non-relativistic case) is also of interest. In
that case both terms in the Eq. (16) need to be retained and to lowest
order in $v_{F}$, $W_{D}$ is given by Eq. (17). Further the missing weight in the universal background below the interband
absorption edge is in the same approximation 
\begin{equation}
\frac{e^{2}\pi}{4h}\sqrt{2m\mu}v_{F}[1-\sqrt{\frac{mv_{F}^{2}}{2\mu}}]
\end{equation}
which does not depend on $\mu$ in the same way as (17). It is clear
that no sum rule applies in this limit as well. The pure relativistic
case is unique and special. 
\begin{figure}[tp]
\begin{centering}
\includegraphics[width=3.1in,height=3in]{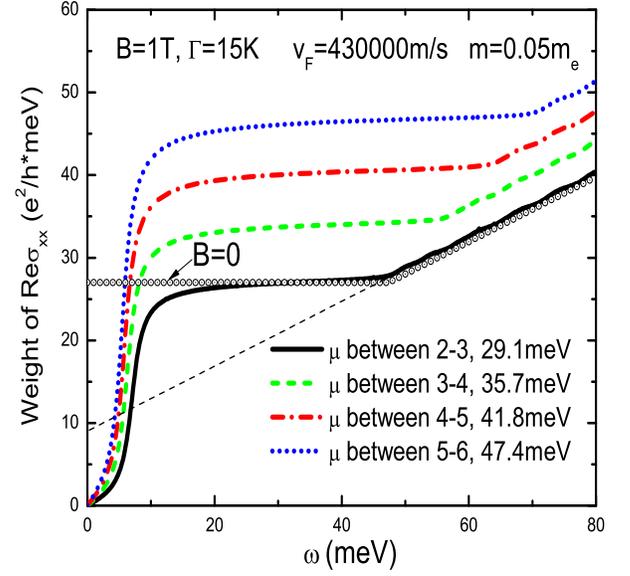} 
\par\end{centering}

\caption{(Color online) The partial optical spectral weight $W(\omega)$ to
an upper cut off $\omega$ (in meV) for four values of chemical potential
$\mu=29.1$ meV (solid black curve), $\mu=35.7$ meV (dashed green
curve), $\mu=41.8$ meV (dash-dotted red curve) and $\mu=47.4$ meV
(dotted blue curve). In all cases the magnetic field is B=1 Tesla,
the Dirac fermi velocity is $4.3\times10^{5}$ m/s and the Schr\"{o}dinger
mass is $0.05m_{e}$. The open circles are for comparison with the
solid black curve but now B is set to zero. The light dashed line
extrapolates the straight line defined above the interband threshold
in the solid black curve. Its intercept with the vertical axis is
a measure of the deviation from pure Dirac behavior brought about
by the presence of a subdominant Schr\"{o}dinger piece in the Hamiltonian
(1).}

\label{fig7} 
\end{figure}

Similar results hold when a magnetic field is present as we can see
in Fig. 7 which gives the total spectral weight below $\omega$ ($W(\omega)$)
under the real part of the longitudinal conductivity $Re\sigma_{xx}(\omega)$
in units of $(e^{2}/h)\times$meV as a function of $\omega$ in meV.
Four cases are shown with $B=1$ Tesla, $\Gamma=15K$, $v_{F}=4.3\times10^{5}$m/s
and the Schr\"{o}dinger mass $m=0.05m_{e}$. The four values of chemical
potential are (solid black curve) $\mu=29.1$meV, (dashed green) $\mu=35.7$meV,
(dashed dotted red) $\mu=41.8$meV and (blue dots) $\mu=47.4$meV.
For the first curve the open circles compare with the $B=0$ case
which starts at a finite value at $\omega=0$ because we have assumed
the clean limit and so the intraband transitions contribute a Dirac
delta function to the frequency dependent conductivity. $W(\omega)$
then remains completely flat until the onset of the interband transitions
set in at $\omega_{\min}=47.8$meV given by Eq. (14). This is quite
a lot smaller than the value of $2\mu\simeq60$meV. The curve for
$B=0$ follows well the solid black curve for $B=1T$ except of course
that in this instance the Drude is replaced by the intraband peak
and so $W(\omega)$ starts from zero at $\omega=0$ and becomes equal
to the black dots only after all the spectral weight in this line
has been picked up. Above the interband onset both curves rise linearly
with increasing $\omega$ with the slope of this line set by the value
of the universal background. If we extrapolate it to $\omega=0$ we
find that it intercepts the $\omega=0$ axis at a non zero value given
by $\frac{e^{2}\pi}{8h}\omega_{\min}(A-1)\equiv\frac{e^{2}\pi}{4h}mv_{F}^{2}(A-1)^{2}$
with\ $A\equiv\sqrt{1+\frac{2\mu}{mv_{F}^{2}}}$. For $m=\infty$,
$A=1$ and this intercept is equal to zero which applies only for
the pure relativistic case. By contrast in topological insulators
$A$ is not one. Should one know independently the value of the chemical
potential one could use the value of the intercept $\frac{e^{2}\pi}{4h}mv_{F}^{2}(A-1)^{2}$
to determine $mv_{F}^{2}$ which is a central parameter in assessing
the deviations from the pure relativistic case expected in a given
topological insulator. Note with reference to the lower frame in Fig.
6 that an extrapolation of the near linear region of the black curve
for $m=m_{e}$, to $\omega=0$ will give almost zero intercept and
this reflect the fact that in this case the Schr\"{o}dinger correction
to pure Dirac is small.

Returning to the data in Fig. 7, $W(\omega)$ rises to a larger magnitude
for the plateau between the saturation of the intraband spectral weight
and the threshold for the interband transition as the chemical potential
is increased. This reflects the fact that as $\mu$ increases, more
interband lines are lost because of Pauli blocking and some of their
optical weight is transferred to the intraband line. The onset of
the interband transition beyond the intraband plateau in $W(\omega)$
increases with increasing $\mu$ but the linear rise above this critical
photon energy has the same slope in all cases. The small variations
about a straight line seen in this region reflect of course the underlying
Landau level structure which is smeared almost completely by the broadening
$\Gamma=15K$. At higher energies all the curves become parallel to
each other but are displaced upward as $\mu$ is increased. In graphene,
as can be seen in Fig. 7 of Ref. [17], the curves would all merge
in this region in contrast to what we find here for topological insulators
when the value of the subdominant Schr\"{o}dinger part to the Hamiltonian
is significant. It is the value of the intercept with the verticle
axis of the straight line variation at large energies which sets the
displacement between these parallel lines and as we have seen graphene
corresponds to the limit of zero intercept.

\section{Summary and Conclusions}

We have calculated the absorptive part of both longitudinal and transverse
(Hall) AC dynamic magneto conductivity for quadratic electronic bands
modified by a small spin orbit coupling. This provides a mechanism
for interband optical transitions between the split helical bands.
These optical transitions are additional to the main intraband (Drude)
absorption which is the only process possible when the spin-orbit
coupling is not present. We find that for the rather small values
of Fermi velocity representative of the semiconductors presently used
in spintronic applications, the interband absorption is small and
that this background is likely not to be seen as distinct from the
main Drude intraband absorption. However as the Fermi velocity $v_{F}$
is increased a distinct interband background emerges which takes on
the same universal values of $\frac{e^{2}\pi}{8h}$ as observed in
grahene except for a degeneracy factor of 4 not applicable in this
case. The universal background however is confined to a definite very
limited range of photon energies with upper and lower cut off dependent
on the value of the chemical potential and also on the magnitude of
the magnetic field when $B$ is also present. Comparison of these
results with those obtained in the opposite limit, when instead the
Schr\"{o}dinger term is a small perturbation on a dominant spin orbit
term, provides new insight into the case of topological insulators.
There, the interband transitions are much more prominent, extend to
large energies and are cut off only by the band edge. For no magnetic
field ($B=0$) we were able to obtain analytic results which are valid
for any value of $m$ and $v_{F}$ and these confirmed aspects of
our finite $B$ results. We showed that the interband background contains
a total optical spectral weight of $\frac{e^{2}\pi}{8h}4mv_{F}^{2}$
which is independent of $\mu$. This spectral weight is small when
either $m$ or $v_{F}$ is small and it is large when both $m$ and
$v_{F}$ are large as in topological insulators. In particular for
$m\rightarrow\infty$ we recover the known result for graphene where
only the Dirac contribution is present, and the interband transitions
extend to the band cut off.

When the chemical potential $\mu$ is increased, a well known result
for graphene is that all the optical spectral weight lost in the interband
transitions is found transferred to the intraband transitions. This
conservation of spectral weight no longer applies when a subdominant
Schr\"{o}dinger piece is added to a pure Dirac Hamiltonian. While these
violations are small for typical parameters associated with present
day known topological insulators, measuring them allows one to determine
the critical parameter $mv_{F}^{2}$ which provides the information
on the expected deviations of optical properties from the pure Dirac
case. In the spintronic limit we find that the amount of spectral
weight in the interband background is almost fixed while at the same
time the intraband contribution increase with increasing chemical
potential.

\begin{acknowledgments} This work was supported by the
Natural Sciences and Engineering Research Council of Canada (NSERC)
and the Canadian Institute for Advanced Research (CIFAR). This material is also based upon 
work supported by the National Science Foundation under the NSF EPSCoR Cooperative 
Agreement No. EPS-1003897 with additional support from the Louisiana Board of Regents.
\end{acknowledgments}

\appendix

\section{Derivation of the Drude spectral weight at zero magnetic field}

The Hamiltonian is given by Eq. (1) and here we write $\alpha=\hbar v_{F}$.
\begin{equation}
H_{0}=\frac{\hbar^{2}k^{2}}{2m}+\alpha(k_{x}\sigma_{y}-k_{y}\sigma_{x})
\end{equation}
The velocity is 
\begin{eqnarray*}
v_{x} & = & \frac{\partial H_{0}}{\partial k_{x}}=\frac{\hbar^{2}k_{x}}{m}+\alpha\sigma_{y}=v_{D}I+\alpha\sigma_{y},\\
v_{y} & = & \frac{\partial H_{0}}{\partial k_{y}}=\frac{\hbar^{2}k_{y}}{m}-\alpha\sigma_{x}=v_{D}I-\alpha\sigma_{x}.
\end{eqnarray*}
The matrix Green's function can be expanded in matrix spectral function
$\hat{A}(\mathbf{k},\omega)$ as 
\begin{equation}
\hat{G}(\mathbf{k},z)=\int_{-\infty}^{\infty}\frac{d\omega}{2\pi}\frac{\hat{A}(\mathbf{k},\omega)}{z-\omega}
\end{equation}
The conductivity $\sigma_{xx}(\omega)$ is given by 
\begin{eqnarray}
 &  & \sigma_{xx}(\omega)=-\frac{e^{2}}{i\omega}\int_{-\infty}^{\infty}\frac{d\omega_{1}}{2\pi}\int_{-\infty}^{\infty}\frac{d\omega_{2}}{2\pi}\frac{[f(\omega_{1})-f(\omega_{2})]}{\omega-\omega_{2}+\omega_{1}+i\delta}\notag\\
 &  & \sum_{\mathbf{k}}Tr\langle v_{x}\widehat{A}(\mathbf{k,}\omega_{1})v_{x}\widehat{A}(\mathbf{k,}\omega_{2})\rangle
\end{eqnarray}
and the real part of the conductivity at zero temperature is given
by 
\begin{eqnarray}
Re\sigma_{xx}(\omega) & = & \frac{e^{2}}{2\omega}\frac{1}{4\pi^{2}}\int_{0}^{k_{cut}}kdkd\theta\int_{-\omega}^{0}\frac{d\omega_{1}}{2\pi}\notag\\
 &  & Tr\langle v_{x}\widehat{A}(\mathbf{k,}\omega_{1})v_{x}\widehat{A}(\mathbf{k,}\omega+\omega_{1})\rangle
\end{eqnarray}
The spectral function is the imaginary part of the Green's function
\begin{equation}
\hat{A}(\mathbf{k},x)=-2Im\hat{G}(\mathbf{k},x)
\end{equation}
The Green's function is given by 
\begin{equation}
\hat{G}_{0}(\mathbf{k},i\omega_{n})=\frac{1}{2}\sum_{s=\pm}(1+s\mathbf{F}_{k}\cdot\mathbf{\sigma})G_{0}(\mathbf{k},s,i\omega_{n})
\end{equation}
where 
\begin{equation}
\mathbf{F}_{k}=\frac{(-k_{y},k_{x},0)}{k}
\end{equation}
and 
\begin{equation}
G_{0}(\mathbf{k},s,i\omega_{n})=\frac{1}{i\omega_{n}+\mu-\frac{\hbar^{2}k^{2}}{2m}-s\alpha k}
\end{equation}
After taking the trace we get 
\begin{eqnarray*}
 &  & Re\sigma_{xx}(\omega)\\
 & = & \frac{e^{2}}{4\omega}\int_{0}^{k_{cut}}kdk\int_{-\omega}^{0}\frac{d\omega_{1}}{2\pi}\\
 &  & 2\pi\alpha^{2}[\delta(\widetilde{\omega_{1}}-\alpha k)\delta(\widetilde{\omega_{2}}-\alpha k)+\delta(\widetilde{\omega_{1}}+\alpha k)\delta(\widetilde{\omega_{2}}+\alpha k)\\
 &  & +\delta(\widetilde{\omega_{1}}+\alpha k)\delta(\widetilde{\omega_{2}}-\alpha k)+\delta(\widetilde{\omega_{1}}-\alpha k)\delta(\widetilde{\omega_{2}}+\alpha k)]\\
 &  & +2\pi\frac{\hbar^{4}k^{2}}{m^{2}}[\delta(\widetilde{\omega_{1}}-\alpha k)\delta(\widetilde{\omega_{2}}-\alpha k)+\delta(\widetilde{\omega_{1}}+\alpha k)\delta(\widetilde{\omega_{2}}+\alpha k)]\\
 &  & +4\pi\alpha\frac{\hbar^{2}k}{m}[\delta(\widetilde{\omega_{1}}-\alpha k)\delta(\widetilde{\omega_{2}}-\alpha k)-\delta(\widetilde{\omega_{1}}+\alpha k)\delta(\widetilde{\omega_{2}}+\alpha k)]
\end{eqnarray*}
where 
\begin{eqnarray}
\widetilde{\omega_{1}} & = & \omega_{1}+\mu-\frac{\hbar^{2}k^{2}}{2m}\\
\widetilde{\omega_{2}} & = & \omega_{1}+\omega+\mu-\frac{\hbar^{2}k^{2}}{2m}
\end{eqnarray}
The term with $\delta(\widetilde{\omega_{1}}+\alpha k)\delta(\widetilde{\omega_{2}}-\alpha k)$
and $\delta(\widetilde{\omega_{1}}-\alpha k)\delta(\widetilde{\omega_{2}}+\alpha k)$
will be zero for topological insulator but nonzero for spintronics.
For topological insulators we have for $\mu>0,\omega_{1}\simeq0,\omega_{1}+\mu-\frac{\hbar^{2}k^{2}}{2m}-\alpha k=0\Rightarrow k=\frac{m}{\hbar^{2}}[-\alpha+\sqrt{\alpha^{2}+2\hbar^{2}/m(\omega_{1}+\mu)}]$,
and 
\begin{eqnarray}
 &  & \int_{0}^{k_{cut}}dk\cdot k\delta(\widetilde{\omega_{1}}-\alpha k)\delta(\widetilde{\omega_{2}}-\alpha k)\nonumber \\
 & = & \frac{m}{\hbar^{2}}\frac{-\alpha+\sqrt{\alpha^{2}+2\hbar^{2}/m(\omega_{1}+\mu)}}{\sqrt{\alpha^{2}+2\hbar^{2}/m(\omega_{1}+\mu)}}\delta(\omega)
\end{eqnarray}
\begin{eqnarray}
 &  & \int_{0}^{k_{cut}}dk\cdot k^{2}\delta(\widetilde{\omega_{1}}-\alpha k)\delta(\widetilde{\omega_{2}}-\alpha k)\nonumber \\
 & = & (\frac{m}{\hbar^{2}})^{2}\frac{(-\alpha+\sqrt{\alpha^{2}+2\hbar^{2}/m(\omega_{1}+\mu)})^{2}}{\sqrt{\alpha^{2}+2\hbar^{2}/m(\omega_{1}+\mu)}}\delta(\omega)
\end{eqnarray}
\begin{eqnarray}
 &  & \int_{0}^{k_{cut}}dk\cdot k^{3}\delta(\widetilde{\omega_{1}}-\alpha k)\delta(\widetilde{\omega_{2}}-\alpha k)\nonumber \\
 & = & (\frac{m}{\hbar^{2}})^{3}\frac{(-\alpha+\sqrt{\alpha^{2}+2\hbar^{2}/m(\omega_{1}+\mu)})^{3}}{\sqrt{\alpha^{2}+2\hbar^{2}/m(\omega_{1}+\mu)}}\delta(\omega)
\end{eqnarray}
For $\mu<0,\omega_{1}\simeq0,\omega_{1}+\mu-\frac{\hbar^{2}k^{2}}{2m}+\alpha k=0\Rightarrow k_{0}=\frac{m}{\hbar^{2}}[\alpha-\sqrt{\alpha^{2}+2\hbar^{2}/m(\omega_{1}+\mu)}].$
Because $k_{cut}=\frac{m\alpha}{\hbar^{2}}$ and the energy at this
point $E_{\min}=\frac{-m\alpha^{2}}{2\hbar^{2}}$. The delta function can be rewritten as $\delta(\widetilde{\omega_{1}}+\alpha k)=\frac{\delta(k-k_{0})}{|f^{\prime}(k_{0})|},$
where $|f^{\prime}(k_{0})|=|-\frac{\hbar^{2}k_{0}}{m}+\alpha|=$ $\sqrt{\alpha^{2}+2\hbar^{2}/m(\omega_{1}+\mu)}$.
Thus 
\begin{eqnarray}
 &  & \int_{0}^{k_{cut}}dk\cdot k\delta(\widetilde{\omega_{1}}+\alpha k)\delta(\widetilde{\omega_{2}}+\alpha k)\nonumber \\
 & = & \frac{m}{\hbar^{2}}\frac{\alpha-\sqrt{\alpha^{2}+2\hbar^{2}/m(\omega_{1}+\mu)}}{\sqrt{\alpha^{2}+2\hbar^{2}/m(\omega_{1}+\mu)}}\delta(\omega)
\end{eqnarray}
\begin{eqnarray}
 &  & \int_{0}^{k_{cut}}dk\cdot k^{2}\delta(\widetilde{\omega_{1}}+\alpha k)\delta(\widetilde{\omega_{2}}+\alpha k)\nonumber \\
 & = & (\frac{m}{\hbar^{2}})^{2}\frac{(\alpha-\sqrt{\alpha^{2}+2\hbar^{2}/m(\omega_{1}+\mu)})^{2}}{\sqrt{\alpha^{2}+2\hbar^{2}/m(\omega_{1}+\mu)}}\delta(\omega)
\end{eqnarray}
\begin{eqnarray}
 &  & \int_{0}^{k_{cut}}dk\cdot k^{3}\delta(\widetilde{\omega_{1}}+\alpha k)\delta(\widetilde{\omega_{2}}+\alpha k)\nonumber \\
 & = & (\frac{m}{\hbar^{2}})^{3}\frac{(\alpha-\sqrt{\alpha^{2}+2\hbar^{2}/m(\omega_{1}+\mu)})^{3}}{\sqrt{\alpha^{2}+2\hbar^{2}/m(\omega_{1}+\mu)}}\delta(\omega)
\end{eqnarray}
and hence 
\begin{eqnarray*}
 &  & \int_{-\omega}^{0}d\omega_{1}\frac{\alpha-\sqrt{\alpha^{2}+2\hbar^{2}/m(\omega_{1}+\mu)}}{\sqrt{\alpha^{2}+2\hbar^{2}/m(\omega_{1}+\mu)}}\\
  & = & \frac{\alpha\omega}{\sqrt{\alpha^{2}+(2\hbar^{2}/m)\mu}}-\omega
\end{eqnarray*}
Similarly 
\begin{eqnarray*}
 &  & \int_{-\omega}^{0}d\omega_{1}\frac{(\alpha-\sqrt{\alpha^{2}+2\hbar^{2}/m(\omega_{1}+\mu)})^{2}}{\sqrt{\alpha^{2}+2\hbar^{2}/m(\omega_{1}+\mu)}}\\
 & = & \frac{(\alpha-\sqrt{\alpha^{2}+(2\hbar^{2}/m)\mu})^{2}\omega}{\sqrt{\alpha^{2}+(2\hbar^{2}/m)\mu}}
\end{eqnarray*}
and 
\begin{eqnarray*}
 &  & \int_{-\omega}^{0}d\omega_{1}\frac{(\alpha-\sqrt{\alpha^{2}+2\hbar^{2}/m(\omega_{1}+\mu)})^{3}}{\sqrt{\alpha^{2}+2\hbar^{2}/m(\omega_{1}+\mu)}}\\
 & = & \frac{(\alpha-\sqrt{\alpha^{2}+(2\hbar^{2}/m)\mu})^{3}\omega}{\sqrt{\alpha^{2}+(2\hbar^{2}/m)\mu}}
\end{eqnarray*}
Thus for $\mu<0,$ 
\begin{equation}
Re\sigma_{xx}(\omega)=\frac{e^{2}m(\alpha-\sqrt{\alpha^{2}+(2\hbar^{2}/m)\mu})\sqrt{\alpha^{2}+(2\hbar^{2}/m)\mu}}{4\hbar^{2}}\delta(\omega)
\end{equation}
and for $\mu>0,$ 
\begin{equation}
Re\sigma_{xx}(\omega)=\frac{e^{2}m(\sqrt{\alpha^{2}+(2\hbar^{2}/m)\mu}-\alpha)\sqrt{\alpha^{2}+(2\hbar^{2}/m)\mu}}{4\hbar^{2}}\delta(\omega)
\end{equation}
For spintronics we can perform similar algebra. 
\end{document}